\documentclass[fleqn,usenatbib]{mnras}
\usepackage{newtxtext,newtxmath,bm}
\usepackage[T1]{fontenc}

\DeclareRobustCommand{\VAN}[3]{#2}
\let\VANthebibliography\thebibliography
\def\thebibliography{\DeclareRobustCommand{\VAN}[3]{##3}\VANthebibliography}
\usepackage{graphicx}	% Including figure files
\usepackage{amsmath}	% Advanced maths commands
\usepackage{subcaption}
\usepackage{xcolor}

\usepackage{physics}

\usepackage{orcidlink}
 
\renewcommand{\vec}[1]{\ensuremath{\bm {#1}}} 
 
\title[Bulge, Disc intrinsic alignments]{Intrinsic alignments of bulges and discs }
 
\author[Y. Jagvaral et al.]{Yesukhei Jagvaral\thanks{E-mail:  yjagvara@andrew.cmu.edu}$^1$\orcidlink{https://orcid.org/0000-0001-7068-7037}, 
Sukhdeep Singh$^1$, 
Rachel Mandelbaum$^1$\orcidlink{0000-0003-2271-1527}
\\
$^{1}$Department of Physics, McWilliams Center for Cosmology, Carnegie Mellon University, Pittsburgh, PA 15213, USA }

\date{Accepted XXX. Received YYY; in original form ZZZ}

\pubyear{2021}

\begin{document}
\label{firstpage}
\pagerange{\pageref{firstpage}--\pageref{lastpage}}
\maketitle

% Abstract of the paper
\begin{abstract}
Galaxies exhibit coherent alignments with local structure in the Universe. This effect, called Intrinsic Alignments (IA), is an important contributor to the systematic uncertainties for wide-field weak lensing surveys. On cosmological distance scales, intrinsic shape alignments have been observed in red galaxies, which are usually bulge-dominated; while blue galaxies, which are mostly disc-dominated, exhibit shape alignments consistent with a null detection. However, disc-dominated galaxies typically consist of two prominent structures: disc and bulge. Since the bulge component has similar properties as elliptical galaxies and is thought to have formed in a similar fashion, naturally one could ask whether the bulge components exhibit similar alignments as ellipticals? In this paper, we investigate how different components of galaxies exhibit IA in the TNG100-1 cosmological hydrodynamical simulation, as well as the dependence of IA on the fraction of stars in rotation-dominated structures at $z=0$. The measurements were controlled for mass differences between the samples. We find that the bulges exhibit significantly higher IA signals, with a nonlinear alignment model amplitude of $A_I = 2.98^{+0.36}_{-0.37}$ compared to the amplitude for the galaxies as a whole (both components), $A_I = 1.13^{+0.37}_{-0.35}$.  The results for bulges are statistically consistent with those for elliptical galaxies, which have $A_I = 3.47^{+0.57}_{-0.57}$. These results highlight the importance of studying galaxy dynamics in order to understand galaxy alignments and their cosmological implications. 
\end{abstract}


\section{Introduction}
 
Deflection of light by matter inhomogeneities between the emitter and the observer produces an effect known as \textit{gravitational lensing} which results in distorted images of the light source \citep{dodelson-book-lensing}.  
The lensing is dubbed \textit{strong}   when  multiple images of the same source, or in rare cases Einstein rings, are produced. However, the images of most objects in the Universe experience minute  distortions en route to the observer in a phenomenon known as \textit{weak} gravitational  lensing. 
The statistical correlations of the lensed shapes produced by coherent structure along the line of sight, known as cosmic shear, is used as a direct probe of large-scale structure, and provides cosmological information on dark energy \citep{Kilbinger}.
Upcoming weak lensing surveys such as the Vera C.\ Rubin Observatory Legacy
Survey of Space and Time (LSST)\footnote{ \url{https://www.lsst.org/} },  Roman Space Telescope\footnote{ \url{https://roman.gsfc.nasa.gov/} } High Latitude Survey (HLS) and Euclid\footnote{ \url{https://www.euclid-ec.org/} } should provide unprecedented precision and constrain cosmological parameters to the percent level \citep{weinberg-dark-energy}.  

However, weak lensing measurements suffer from a number of systematic biases that must be corrected for precisely in order to unlock this cosmological constraining power. One important contributor is intrinsic alignments (IA), the tendency of galaxies to coherently align with the density field and produce correlations among  galaxy shapes \citep[see, e.g.,][for a review]{troxel-ishak}. This effect can masquerade as a weak lensing signal and bias the constraints  on cosmological parameters such as the equation of state of dark energy.
Thus, for precision cosmological measurements this effect has to be taken into account either by mitigation or marginalization \citep{weak-lens}. On the flip side, IA are of interest because they can give us insight into galaxy formation and evolution, and into the large-scale structure, since IA is produced by the gravitational interplay between galaxies and the underlying large-scale structure.

Early analytic models for galaxy intrinsic alignments included the linear alignment model \citep{catelan, hirata-seljak}.  
Later extensions included  non-linear contributions \citep{bridle-king,blazek-tatt}. These models can effectively describe the large-scale alignment behavior of elliptical galaxies; however for intermediate and small scales, these models tend to underestimate the alignment.  
In order to address this problem, halo models of IA \citep{schneider-bridle,fortuna} were developed to describe the small scale effects. These analytic models usually depend on assumptions relating the galaxy and the host dark matter halo orientations which may or may not be valid.  
On another frontier, N-body cosmological simulations containing only dark matter (DM) have also been used to study IA by ``painting'' galaxies onto the DM haloes.  
However, the predicted IA signals proved to be entirely dependent on the model used \citep{heymans-2006}. 
 Recently, with the advent of large-volume hydrodynamic simulations, direct study of large-scale IA within  simulations has been enabled.  
 There have been numerous studies of IA within the Illustris, Illustris-TNG, MassiveBlack-II, Horizon-AGN and EAGLE simulations \citep[e.g.,][]{tenneti-ia, eagle-ia,tenneti-disc-ellip,chisari-horizon-ia,samuroff-2020,hill-eagle-1,hill-eagle, zjupa-ia} with a goal of developing accurate models to incorporate into future survey analysis pipelines and to provide insight into how galaxy formation/evolution processes influence the alignment signals.  While such IA measurements will still depend on the subgrid galaxy formation and evolution physics models used in the hydrodynamical simulations, compared with `galaxy painting' models, hydro simulations are more complex and try to capture more aspects of the baryonic physics.

Not all galaxies exhibit similar IA; many studies show that large-scale IA depend on the galaxy location within its dark matter halo (satellite vs central), color (red vs blue), and luminosity  \citep{ia-review1}. 
In observational studies of large-scale IA, galaxy color typically serves as a proxy for morphology and the samples are split by color \citep{ia-review2}.
Red galaxies (which are usually elliptical) are dynamically dispersion-dominated and are believed to exhibit coherent aligned with the large-scale tidal field. This model would explain the measured IA effects for red galaxies \citep{mandelbaum-2006, hirata-2007,okumura-2009, joachimi-2011,singh-2015}.  
In contrast, the dynamics of blue galaxies (which are usually disc-dominated)  are determined by their angular momentum, and 
are influenced by the small scale tidal fields via torque, leading to a correlation in their orientations \citep{hirata-seljak}. 
As a result, the common understanding is that blue galaxies tend to have spin alignments with the cosmic web \citep{spin-alignment-cosmic-web}.  
In observational studies, large-scale IA of galaxy shapes has not been detected for   blue galaxies  so far \citep{mandelbaum-2011,samuroff-2019,johnston-2019}.

In simulations, all of the phase space data are available, making dynamical morphological classification of galaxies possible. \cite{tenneti-disc-ellip} studied dynamically classified disc and elliptical galaxies in MBII and Illlustris, and found a similar trend that early-type galaxies are more strongly aligned than late-types. Later,  \cite{chisari-horizon-ia} came to the  conclusion that spheroidal galaxies tend to be aligned radially towards over-densities and other spheroidals, whereas disc galaxies tend to be oriented tangentially around spheroidals in 3D for the Horizon AGN simulation.  
 Additionally, \cite{shao-2016} studied IA in EAGLE and cosmo-OWLS simulations and reported that  dispersion-dominated (spheroidal) centrals have
a stronger alignment than rotation-dominated (disc) centrals with both their dark matter halo and the distribution of their satellites.

 Most disc galaxies consist of two prominent structures: bulge and disc. Naively, one may assume that bulges are scaled down ellipticals. However, there are some differences. 
 Galactic bulges are further broken down into  classical bulges and disky-bulges (pseudobulges).  Ellipticals and classical bulges are hypothesized to be the products of major galaxy mergers. On the other hand,  pseudobulges are hypothesized to be the product of secular evolution of galactic discs \citep{bulge-book}. Also, \cite{gaditti-2009} concludes that  classical bulges and ellipticals follow offset mass–size relations, suggesting that high-mass bulges cannot be considered as high-mass ellipticals that happen to be surrounded by a
disc.
 Despite the differences between ellipticals and bulges in general, they have similar stellar kinematics and internal properties such as low gas content, low star formation rate and red color.
 Since bulges are thought to form and evolve in a similar fashion to elliptical galaxies and share the same properties,  we investigate how different components of galaxies exhibit IA in a hydrodynamical simulation. 
Although disc galaxy samples have tended to produce null large-scale IA signals (with substantial errors in some cases), it does not mean future surveys can safely ignore them. This paper aims to provide a better understanding of  
the morphological dependence of IA.

 Also, there is some evidence that a simple split of red and blue galaxies is not the full story.  For example, \cite{johnston-2019} reports that the modelling of red/blue-split galaxy alignments may be insufficient to describe samples with variable central/satellite galaxy fractions; \cite{G19} came to a similar conclusion.
 Furthermore, S0 galaxies, which appear red in color, usually have disc-like morphology \citep{vanderKruit-2011}; also, there are low mass blue elliptical galaxies. Both of these factors complicate the color-morphology relation.  IA is a dynamical effect and has no direct physical relationship with galaxy color and chemical composition, with any correlations between IA and color presumably being inherited via the pronounced color-morphology correlation. Therefore, our study focusing on IAs of dynamically classified galaxies will provide better insight. Additionally, since these intrinsic alignment mechanisms are inevitably tied with the environment and the formation of galaxies in their host halo, IA also provides valuable insight into the large-scale structure and galaxy formation/evolution.  
 
 In this work, we do not consider additional structures such as bars, rings, stellar halos and spiral arms. We use \textit{projected}  
 shape correlations of galaxies that are directly comparable and relatable to observed galaxy shape correlations,  instead of IA estimators that rely on a knowledge of the 3D orientation of the galaxies \citep[e.g.,][]{tenneti-ia,eagle-ia}.  

We begin in \S~\ref{theory_sec} by describing the theory and the IA model we have used. In \S~\ref{sim-section} we describe the simulation suite and the analysis methods that we use to quantify IA, including galaxy shape estimators, two-point statistics and misalignment angles.
Next, in \S~\ref{results} we present the measured shapes and  
discuss the various misalignment angles we have investigated, and interpret the measured two-point correlation functions. 
Finally, in \S~\ref{conc} we discuss and conclude our paper.

\section{Theory}
\label{theory_sec}
 
 Weak lensing measurements typically utilize the correlation function of the ellipticity measurements
of many galaxies in order to achieve the statistical power to probe the lensing-induced galaxy shape correlations, compared to the more dominant shape noise \citep{dodelson-book-lensing}.  
The galaxy shears can be modeled as a random term plus two coherent shears. The two coherent shears can be written as $\gamma = \gamma^G + \gamma^I$, where $\gamma^G$ is due to gravitational lensing and $\gamma^I$ is due to intrinsic alignments. Then, the two-point correlation function is 
\begin{equation}
    \langle \gamma \gamma\rangle = \langle \gamma^G \gamma^G\rangle + \langle \gamma^G \gamma^I \rangle  +\langle \gamma^I \gamma^G \rangle   +\langle \gamma^I \gamma^I \rangle% = \xi_{GG} + \xi_{GI}+\xi_{IG}+\xi_{II}
\end{equation}
 
The first term, $\langle \gamma^G \gamma^G\rangle$ is the desired signal in weak lensing surveys. For completely random shape orientations in the absence of lensing,  
 the  terms  $\langle \gamma^G \gamma^I\rangle$ and $\langle \gamma^I \gamma^I\rangle$, average to zero.   
 In reality, due to coherent alignments of galaxy shapes with the large-scale structure, these two terms acquire non-zero values, thus contaminating the weak lensing signal. Intuitively, one can think of the $\langle \gamma^G \gamma^I\rangle$ and the  $\langle \gamma^I \gamma^G\rangle$ terms as describing a pair of galaxies, where one in the foreground (via its tidal field) lenses the one in the background  
whilst  the same tidal field (of the foreground galaxy) also affects its own shape and that of nearby galaxies.  In contrast, the $\langle \gamma^I \gamma^I\rangle$ term reflects a scenario where two nearby galaxies   are affected by
the same tidal field \citep{hirata-seljak,troxel-ishak,Singh-2016}.

\subsection{ The nonlinear alignment (NLA) model}
 
In this paper, we will interpret the intrinsic alignment two-point correlation functions measured in simulations in terms of the nonlinear alignment model.  This subsection describes how we calculate the theoretical predictions for this model. We will start with the linear alignment (LA) model, and then the non-linear alignment (NLA) model that incorporates the non-linear power spectrum will be briefly discussed.
The LA model assumes that IA are determined by the tidal field
at the time of galaxy formation \citep{catelan}. Consequently we can write the intrinsic shear in terms of the primordial potential $\phi_p$:
\begin{equation}
\gamma^I =  (\gamma^I_+,\gamma^I_-) =- \frac{C_1}{4\pi G} (\partial^2_x - \partial^2_y, \partial_x\partial_y) \phi_p
\end{equation}
where $C_1$ is the alignment amplitude and $(x,y)$ forms an arbitrary coordinate system in the plane of the sky.  
Here our sign convention is the same as in \cite{singh-2015}. Next,  assuming a linear galaxy bias between matter
overdensities and galaxy densities $\delta_g = b \delta_m$, the power
spectrum of galaxy-shape and shape-shape correlations are as follows \citep{hirata-seljak}:

\begin{align}
			P_{g+}(\vec{k})&=A_I b \frac{C_1\rho_{\text{crit}}\Omega_m}{D(z)} \frac{k_x^2-
			k_y^2}{k^2} P_\delta^\text{lin} (\vec{k})\label{eqn:Pg+}\\
			P_{++}(\vec{k})&=\left(A_I \frac{C_1\rho_{\text{crit}}\Omega_m}{D(z)} 
			\frac{k_x^2-k_y^2}{k^2} \right)^2P_\delta^\text{lin} (\vec{k})\label{eqn:P++}\\
			P_{g\times}(\vec{k})&=A_I b\frac{C_1\rho_{\text{crit}}\Omega_m}{D(z)}\frac{k_x 
			k_y}{k^2}P_\delta^\text{lin}(\vec{k}).\label{eqn:Pgx}
		\end{align} 
 
		Here $P_\delta^\text{lin}$ is the linear matter power spectrum, while $P_{g+}$  is the
        cross-power spectrum between the galaxy density field and the 
		shear component along the line joining the galaxy pair.  
		$P_{++}$ 
		is the shape-shape correlation of the shear component along the line joining the galaxy
        pair.  $P_{g\times}$ is the
        cross-power spectrum between the galaxy density field and the 
		shear component at  45$^{\circ}$   from the  line joining the galaxy pair.
        Following the convention in the  literature \citep{joachimi-2011,singh-2015}, we fix $C_1\rho_\text{crit}=0.0134$ and use  the $A_I$ parameter to quantify the IA amplitude.   
The NLA model incorporates the non-linear matter power spectrum  to partially describe 
the non-linear regime \citep{bridle-king}, so  the non-linear
matter power spectrum  ($P_\delta^\text{nl}(\vec{k})$) replaces the linear power spectrum ($P_\delta^\text{lin}(\vec{k})$) in Eqs.~\eqref{eqn:Pg+}--\eqref{eqn:Pgx}.
 
\subsection{Modeling the real-space observables}
To measure intrinsic alignments in data and simulations, we typically use correlation functions, which can be obtained by fourier transforming the power spectra as
\begin{equation}\label{eq:xi_to_w}
    \xi_{ab} (\vec{r}) = \int d^3\vec{k} P_{ab}(\vec{k})e^{i\vec{k}\cdot\vec{r}}
\end{equation}
where the indices $a,b\in(g,+)$ represent the two different fields being correlated (in an auto correlation, $a=b$), and $r$ is the 3D separation for a pair of galaxies. Two different samples of galaxies are cross-correlated in order for one sample to trace the biased matter density ($g$) and the other to trace the intrinsic shear ($+$).
 
We further integrate the correlation functions along the line-of-sight to obtain the 2D projected correlation functions as \citep[see e.g. ][ for more details]{singh-2015}:
\begin{equation}\label{eq:xi_to_w}
w_{ab} (r_\mathrm{p}) = \int_{-\Pi_\mathrm{max}}^{\Pi_\mathrm{max}} \mathrm{d}\Pi\, \xi_{ab} (r_\mathrm{p}, \Pi).
\end{equation}
$\Pi$	 is the
line-of-sight separation and $r_\mathrm{p}$ is the projected separation for a pair of galaxies, with $r^2=r_p^2+\Pi^2$. 
 
For faster and numerically stable calculations of the theoretical model, we follow the method of \cite{Singh-2021}, where the projected correlation function from the previous equation is written as 
    \begin{equation}		
        w_{ab}(r_p)=\sum_{\ell=0}^2 2		\int_0^{\Pi_\text{max}}d\Pi\, \xi_{ab,2\ell}(r) L_{2\ell}\left(\frac{\Pi}{r}\right).
				\label{eq:w_model}
	\end{equation}
 
	 where $L_{2\ell}$ are the Legendre polynomials of order $2\ell$, and $\xi_{ab,2\ell}(r)$ are the multipoles of the correlation function obtained from the power spectra as 
	\begin{equation}
				\xi_{ab,2\ell}(r)=(-1)^\ell\alpha_{2\ell}(\beta_a,\beta_b)\frac{1}{2\pi^2}\int dk k^2 P_{ab}(k)j_{2\ell}(kr).
				\label{eq:xi_multipole_model}
			\end{equation}
 
    Here $j_{2\ell}$ denotes the spherical Bessel functions of order $2\ell$, $P_{ab}$ refers to the cross power spectra as defined in Eq.~\eqref{eqn:Pgx} and the galaxy power spectrum for a sample with galaxy bias $b$ is $P^\mathrm{nl}_{gg}=b^2 P^\mathrm{nl}_{\delta}$.  
    We use the FFTlog implementation in  the {\sc MCFIT} package \citep{Li-2019} to obtain  $\xi_{ab,2\ell}$. The prefactors $\alpha_{2\ell}(\beta_a,\beta_b)$ are as follows:
    \begin{align}
    \begin{split}
				\alpha_0(\beta_a,\beta_b)&=1+\frac{1}{3}(\beta_a+\beta_b)+\frac{1}{5}\beta_a\beta_b\label{eq:rsd_alph0}\\
				\alpha_2(\beta_a,\beta_b)&=\frac{2}{3}(\beta_a+\beta_b)+\frac{4}{7}\beta_a\beta_b\\%\label{eq:rsd_alph2}\\
				\alpha_4(\beta_a,\beta_b)&=\frac{8}{35}\beta_a\beta_b%\label{eq:rsd_alph4}
	\end{split}
	\end{align}
 
	The $\beta$ factors model the redshift space anisotropy of the correlation function ($\beta=f/b$ in redshift space distortion measurements with $f$ being the rate of growth of structure.  
	Since we use the simulated data in real space rather than redshift space, $\beta_g=0$. The IA model, on the other hand, has an additional anisotropy from the use of projected shapes \citep{Singh-2016}, such that $\beta_+=-1$.
 
 These projected two point functions $w_{g+}$ and $w_{++}$ are widely used in observational IA studies.
  
 Together with $w_{gg}$, these equations can be used to model the IA in the simulations and derive IA model parameters that can be directly compared with observational studies.

\section{The Simulation and Analysis Methods}\label{sim-section}
 Here we describe the simulated data and explain the methods we use to measure simulated galaxy and DM halo shapes, misalignment angles, and two point statistics, and to decompose/classify galaxies dynamically.
 
\subsection{Simulated data}\label{simulation}
 
Here, we succinctly introduce the IllustrisTNG simulation used in this work \citep[for more information, please refer to][]{ tng-bimodal,pillepich2018illustristng, Springel2017illustristng, Naiman2018illustristng, Marinacci2017illustristng,tng-publicdata}.
The IllustrisTNG100-1 is a large-volume hydrodynamical simulation with a box side length of 75 Mpc/h. The simulation was run using the moving-mesh code Arepo \citep{arepo} and  has $2 \times 1820^3$ resolution elements with a gravitational softening length of 0.7 kpc/h 
for dark matter and star particles, respectively. 
The dark matter particle mass is $7.46\times 10^6  M_\odot$ and star particle masses are variable. The model in the simulations for galaxy formation and evolution includes radiative gas cooling and heating; star formation in the ISM; stellar evolution with metal enrichment from supernovae; stellar, AGN and blackhole feedback; formation and accretion of supermassive blackholes \citep{tng-methods}.
The DM halos within the simulation were cataloged using friends-of-friends (FoF) 
methods \citep{fof}, and the subhalos  
were cataloged using the SUBFIND algorithm \citep{subfind}. 
The simulation suite includes 100 snapshots at different redshifts; we use the latest snapshot at $z=0$ for our analysis.    
 \subsection{Shapes of Halos and Galaxies}\label{shapes_methods}
 
To measure the shapes of galaxies and DM halos we utilize the mass quadrupole moments (often incorrectly referred to as the inertia tensor 
).  We use three different definitions of these moments -- ${ I}_{ij}$ (simple), $\hat{ I}_{ij}$ (reduced) and $\widetilde{ I}_{ij}$  
(reduced iterative) -- with the first two defined as:
\begin{equation}
\label{eq:simple_IT}
{ I}_{ij} = \frac{\sum_n m_n {r_{ni} r_{nj}} }{\sum_n m_n}
\end{equation}
\begin{equation}
\label{eq:reduc_IT}
\hat{ I}_{ij} = \frac{\sum_n  \frac{m_n} {r_{n}^2} r_{ni} r_{nj}}{\sum_n  \frac{m_n} {r_{n}^2} }.
\end{equation}
Here the summation index $n$ runs over all  particles of a given type  
in a given galaxy, where  $m_n$ is the mass of the $n^{\rm th}$ particle and
\begin{equation}\label{eq:rad_dist}
r_{n}^2 = \sum_{i=1}^3 r_{n,i}^2 = x_{n}^2 + y_{n}^2 +z_{n}^2
\end{equation}
is the distance between the centre of mass and the $n^{\rm th}$ particle, with $i$ indexing the three spatial directions\footnote{We assign weights of 1 to the particles with $|r|< 10^{-3}$ kpc$/h$ for numerical stability. We have tested how much of an effect this weight replacement has on the shape measurements compared to replacing only the weights for particles with strictly r=0, and see differences of about 20-30\% consistent with random scatter.}. 

The reduced mass quadrupole moment in Eq.~\eqref{eq:reduc_IT} upweights particles that are closer to the centre of the halo/galaxy, thus downweighting the loosely bound particles   in the outer regions. However, this specific moment definition imposes a spherical symmetry on the halo/galaxy, and produces rounded shapes compared to the simple mass quadrupole moment and compared to the true shape of the halo/glaxy. 

To devise an alternate estimator that upweights particles at small separations without producing a bias, an  iterative procedure is used to obtain $\widetilde{I}_{ij}$:  after the initial calculation of the $\hat{ I}$, the particles are rotated so that the three unit eigenvectors of the mass quadrupole moment, defined   as $\vb s_\mu=\{s_{x,\mu},s_{y,\mu},s_{z,\mu}\}^\tau$ and $\mu \in \{a,b,c\}$, 
 
are aligned with the x-, y-, z-axes, respectively. The half-lengths of the principal axes of the ellipsoid are given by $a\propto\sqrt{\omega_a}$, $b\propto\sqrt{\omega_b}$, and $c\propto\sqrt{\omega_c}$, such that $a \geq b \geq c$ and $\omega_a, \omega_b ,\omega_c$ are the   eigenvalues of the mass quadrupole moment.
Then, the radial distance, $r_{n}^2$ in Eq.~\eqref{eq:rad_dist}, is replaced with the elliptical radial distance:
\begin{equation}
\tilde{r}_{n}^2 = \left(\frac{x_{n}}{a}\right)^2 +
\left(\frac{y_{n}}{b}\right)^2 +
\left(\frac{z_{n}}{c}\right)^2.
\end{equation}
 
Then, $\widetilde{I}_{ij}$ is calculated using $\tilde{r}_{n}^2$ as:
\begin{equation}
\label{eq:iter_IT}
\widetilde{ I}_{ij} = \frac{\sum_n  \frac{m_n} {\tilde{r}_{n}^2} r_{n,i} r_{n,j}}{\sum_n  \frac{m_n} {\tilde{r}_{n}^2} }.
\end{equation}
The iteration continues until the change in eigenvalues to the next step is less than 1  percent \citep{iterative}.
  
To predict the projected alignment signals, we need to use the 3D mass quadrupole moments to define 2D projected shapes. Following \cite{joachimi-2013}, we can obtain the projected 2D ellipse by solving $\bm{x}^\tau {\mathbf W}^{-1} \bm{x} = 1$, where 
\begin{equation}
\label{eq:ellipseprojection1}
{\mathbf W}^{-1} = \sum_{\mu=1}^3 \frac{ \bm{s}_{\perp,\mu} \vb \bm{s}_{\perp,\mu}^\tau}{\omega_\mu^2} - \frac{\bm{k} \bm{k}^\tau}{\alpha^2}\;,
\end{equation}
and
\begin{equation}
\label{eq:ellipseprojection2}
\vb k = \sum_{\mu=1}^3 \frac{s_{\parallel,\mu} \bm{s}_{\perp,\mu}}{\omega_\mu^2}\;~~~\mbox{and}~~
\alpha^2 = \sum_{\mu=1}^3 \left( \frac{s_{\parallel,\mu}}{\omega_\mu} \right)^2\;.
\end{equation}
Here, $\vb s_{\perp,\mu} = \{s_{x,\mu},s_{y,\mu}\}^\tau$ are the eigenvectors projected along the projection axis (for which we arbitrarily choose the z-axis of the 3D simulation box).  

Then, the  two components of the galaxy ellipticity  can be expressed in terms of the symmetric tensor ${\mathbf W}$ 
\begin{equation}
(e_1,e_2) = \frac{(W_{xx}-W_{yy}, 2W_{xy})}{W_{xx} + W_{yy} + 2 \sqrt{\mathrm{det}\mathbf{W}}}.
\end{equation}
 For the special case that the $\bm{s}_c$ lies perfectly along the projection axis, the absolute value of the ellipticity is $|e| = (a-b)/(a+b)$. In terms of the projected simulation box, the $x,y$ directions correspond to the positive and negative direction of $e_1$ (since we projected along the $z$ direction).

\subsection{Misalignment angle}
 
  DM halos are usually modeled by assuming the  collisionless cold dark matter particles gravitationally collapse in an ellipsoidal shape \citep{sheth-ellipsoidal-collapse}.   
  However, the galaxies themselves exhibit a variety of shapes ranging from very oblate thin discs to ellipticals. This shape variety among galaxy populations  is hypothesized to be a consequence of angular momentum redistribution during galaxy formation and evolution \citep{vanderKruit-2011}. 
  Therefore, the study of how galaxy shapes and angular momenta align with those of their host DM halo 
will provide insight into the astrophysical processes that guide   galaxy evolution.
The orientation of a galaxy relative to its host DM halo can be described by three misalignment angles defined as
\begin{equation}
\theta_{\mu \lambda} = \cos^{-1} ( \bm{s}^\mathrm{dm}_\mu \cdot  \bm{s}^\mathrm{g}_\lambda)
\end{equation}
where $\lambda, \mu=a,b,c$,  
and   $\bm{s}^\mathrm{dm}_\mu  (\bm{s}^\mathrm{g}_\mu)$ is the unit vector defining one of the principal axes of the DM halo (galaxy). In 
addition, we calculate the misalignment angle of the minor axis $\mathbf{s}^\mathrm{g}_c$ with the angular momentum of the DM halo and of the galaxy:
\begin{equation}
\theta_{\bm{L}c} = \cos^{-1} ( \bm{L}^\mathrm{k} \cdot  \bm{s}^\mathrm{g}_c)
\end{equation}
 
where  
\begin{equation}
\bm{L}^\mathrm{k} =  \sum_{n=1}^{N} m^{\mathrm{k}}_n \bm{x}_n \cross \bm{v}_n.  
\end{equation}
 
Here $\mathrm{k} \in \text{(dm, gal)}$, $m^{\mathrm{k}}_n$ is the mass of particle $n$, $\bm{x}_n$ is its position, and $\bm{v}_n$ is its velocity relative to the origin of the subhalo/galaxy.  
The origin of the subhalo/galaxy 
is taken as the position of the most bound particle in the subhalo/galaxy.  

\subsection{Two-point correlation function estimators}\label{subsec:2pcf-est}

 We measure the galaxy two-point correlation functions using the standard Landy-Szalay estimator \citep{landy}, 
 %\sukhdeep{I rewrote this}\response{ok}
\begin{equation}
\xi_{gg}(r_{\rm p}, \Pi) = \frac{D D - DR - DR + RR}{RR},
\end{equation}
\noindent
where $DD$, $RR$ and $DR$ are weighted counts of galaxy-galaxy, random-random and galaxy-random pairs, binned based on their perpendicular and line-of-sight separation, $r_p$ and $\Pi$ \footnote{All of the two-point statistic were measured using the  \href{https://halotools.readthedocs.io/en/latest/}{HALOTOOLS} package v0.7 \citep{halotools} and the supporting \href{https://github.com/duncandc/halotools_ia}{halotools\_ia} package .}.  

The cross correlation function of galaxy positions and intrinsic ellipticities, $\xi_{g+}(r_\mathrm{p}, \Pi)$, can be similarly measured using a modified Landy-Szalay estimator \citep{mandelbaum-2011}  
as a function of $r_\mathrm{p}$ and $\Pi$:
\begin{equation}
\xi_{g+} (r_\mathrm{p}, \Pi) = \frac{S_+D - S_+R}{RR}.
\end{equation}
Similarly, the shape-shape correlation can be estimated as 
\begin{equation}
\xi_{++} (r_\mathrm{p}, \Pi) = \frac{S_+S_+}{RR}.
\end{equation}
Here 
\begin{equation}
S_+D \equiv \frac{1}{2} \sum_{\alpha\neq \beta} w_\alpha w_\beta \, e_{+}(\beta|\alpha),
\end{equation}
\begin{equation}
S_+S_+ \equiv \frac{1}{4} \sum_{\alpha\neq \beta}    w_\alpha w_\beta \,  e_{+}(\alpha|\beta) e_{+}(\beta|\alpha),
\end{equation}
 represent the shape correlations,  
 where  $e_{+}(\beta|\alpha)$ is the $+$ component of the ellipticity of galaxy $\beta$ (from the shape sample) measured relative to the direction of galaxy $\alpha$ (from the density tracer sample) and $w_\alpha$ ($w_\beta)$ are weights associated with galaxy $\alpha$ ($\beta$).

 Finally, the projected two-point correlation functions (Eq.~\ref{eq:xi_to_w}) are approximated as sums over   the line-of-sight separation ($\Pi$) bins:
 \begin{equation}\label{eq:xi_to_w_sum}
w_{ab} (r_\mathrm{p}) = \sum_{-\Pi_\mathrm{max}}^{\Pi_\mathrm{max}} \Delta\Pi \,\xi_{ab} (r_\mathrm{p}, \Pi),
\end{equation}
where  $a,b\in(g,+)$. 

  Additionally, in both the model and the measurement we employ a $\Pi_\text{max}$ value of 20 Mpc, and the model has a $k_\text{min}$ cutoff of $\pi/L_\text{box} = 0.4$ h/Mpc. These choices are made  in order to exclude the modes that are not captured by the finite size of the simulation box.   
 We fit the three correlation functions ($w_{gg}$, $w_{g+}$,
and $w_{++}$) jointly   to the NLA model, with the galaxy
bias $b$ and the intrinsic alignment amplitude AI as the fit parameters. The same set of density tracers, including all galaxies above $\log_{10}(M_*/M_\odot) =10 $,  were used for all correlation functions and fits.
For our fits, we have used an analytic estimate of the covariance matrix derived from the dominant Gaussian contribution; under this assumption, the covariance matrix includes a noise term and cosmic variance (for more details see \citealt{singh-covarience,samuroff-2020}). 
 
\subsection{2D kinematic decomposition model } \label{2d-method}
 
In this subsection, we describe the dynamical model that probabilistically assigns each particle to either the bulge or the disc component of the galaxy (for more details, see \citealt{gal_decomp}). 
The model identifies the two galaxy components through two physically-motivated assumptions:
\begin{enumerate}
  \item Disc stars' angular momentum is approximately aligned with the total angular momentum of the galaxy, while the orientation of  bulge stars' angular momentum is randomly distributed.
  \item Disc stars' orbits are approximately circular, while bulge stars' orbits are elongated or circular.  
\end{enumerate}
 
To implement a dynamical decomposition based on those two principles, we define the following two parameters of interest:
\begin{itemize}

  \item  $j_\text{r} \equiv \frac{j_\text{star}} {j_\text{circ}(r)}$, where $\bm{j}_\text{star}$ is the angular momentum of a single star particle and $j_\text{star}$ is its magnitude;   $j_\text{circ}(r) = r\, v_\text{circ}(r) = r \, \sqrt{\frac{GM(r)}{r}}$   is the expected angular momentum for a circular orbit at the same position as that star,
where $M(r)$ is the total mass (across all types of particles -- stars, gas, dark matter) contained within that radius.  This parameter indicates the type of orbit taken by the star particle. Stars on circular orbits will have $j_\text{r} \sim  1$, while those on elliptical orbits can have $j_\text{r}$ either above or below 1. 
 
  \item $\cos\alpha$ is the cosine of the angle between the angular momentum vector of the star particle and the total angular momentum of the galaxy. A concentration of particles at $\cos\alpha \sim 1$  signals that the galaxy contains a disc structure, since particles preferentially have their angular momentum aligned with that of the galaxy overall. Spread within the  angular parameter  $\cos\alpha$  signifies the disordered motion among bulge stars. 
  \end{itemize}
 
 To build this 2D model, we assume bulge stars should exhibit a flat distribution in  $\cos(\alpha)$, so their distribution in this plane should be solely dependent on  $j_{r}$.  
 On the other hand, we assume the distribution of disc stars depends on both parameters. 
Hence, we consider the following model for the probability distribution of star particles:
\begin{equation}%\label{1d_eq}
   p_\text{star}(j_\text{r},\cos\alpha) \equiv (1-f^\text{disc}) \, p_\text{bulge} (j_\text{r},\cos\alpha) + f^\text{disc} \, p_\text{disc}  (j_\text{r} , \cos\alpha).
\end{equation}
Here $p_\text{bulge}$ and $p_\text{disc}$ are the probability distributions (both normalized to integrate to 1)  that a star at a given point in this 2D space belongs to the bulge or the disc.   We assume independent distributions for the two parameters in $p_\text{bulge}$, parametrizing it as:
\begin{equation}
p_\text{bulge} (j_\text{r},\cos\alpha) = \Gamma (j_\text{r}) \cdot U(\cos\alpha), 
\end{equation}
where $\Gamma(j_\text{r})$ is the Gamma  distribution and $U(\cos\alpha)$ is the Uniform distribution on the interval [-1,1]. The Gamma   
distribution was chosen empirically to fit the skewed distribution in $j_\text{r}$. In contrast, for $p_\text{disc}$ we have used a non-parametric representation via kernel density estimation.  

    The 2D model is built deterministically, after which we can generate Monte Carlo realizations of the model, assigning star particles to the bulge or to the disc. In this work, we will focus on a single realization of the Monte Carlo simulation; the probabilistic nature of the Monte Carlo method was explored in \cite{gal_decomp}.  
  Additionally, that paper demonstrated the improved robustness of our kinematic decomposition compared to a widely used method in the literature involving cuts on the \textit{circularity} parameter. The number fraction of disc-dominated galaxies at a given stellar mass obtained by the model agrees well with observations for masses exceeding $\log_{10}(M_*/M_\odot)=10$. Furthermore, we showed that the S\'ersic indices and half-mass radii for the bulge and disc components agree well with those of real galaxies from SDSS and CANDELS. There were some shortcomings as well, such as that the galaxies classified as disc-dominated contained a significant number of red galaxies alongside the blue population.  However, this result had been seen before for IllustrisTNG \citep{tng-bimodal}, implying it is a feature of the galaxy formation model in the simulation rather than a feature of our kinematic decomposition.
Given the successes of the model, our dynamical morphological classification and the results derived from using these classifications are robust. However, the color classifications and the results derived from them may not be as robust as the dynamical classification.

\subsection{Sample selection}\label{sample_sel}

We employ a minimum stellar mass threshold of $ \log_{10}(M_*/M_\odot) =10 $ for all galaxies, using their stellar mass from  the SUBFIND catalog.  A previous study \citep{gal_decomp} motivates the mass cut on the shape sample, where we concluded that the   disc fractions from TNG100-1 agree well with observational values down to $ \log_{10}(M_*/M_\odot) =10 $, below which  we believe the resolution of the simulation may be affecting the results.   
 Also, we have checked that using a lower mass on the density tracers, to include more galaxies, did not change our $A_I$ amplitudes.
 
 Based on the model discussed in the previous section, we classify galaxies into three bins in $f_\mathrm{disc}$, each containing the same number of galaxies,  
  where $f_\mathrm{disc}$ is  the fraction of stellar mass in the disc for a given galaxy. 
 For the rest of this paper, we will refer to the bin with the lowest mean $\langle f_\mathrm{disc} \rangle = 0.37$ as the \textit{Elliptical} sample, the bin with the highest $\langle f_\mathrm{disc} \rangle = 0.83$ as the \textit{Pure Disc} sample, and the middle bin with $\langle f_\mathrm{disc} \rangle = 0.68$ as the two-component \textit{Disc+Bulge} sample. Further, the calculations for the \textit{Disc+Bulge} sample are carried out separately for the two components of each galaxy: the \textit{Bulge (only)} sample and the \textit{Disc (only)} sample.  
 
 Often galaxies are split into two groups based on morphology, but in this study we chose to split the sample into three bins in order to gain more insight into how IA depends on $f_\mathrm{disc}$.  
 We chose the sample names for simplicity, but should emphasize that the galaxies in the \textit{Pure Disc} sample contain a non-negligible fraction of dispersion-dominated stars.  Conversely, the galaxies in the \textit{Elliptical} sample contain a non-negligible fraction of rotation-dominated stars. These  non-negligible but small fractions do not lend themselves to making reliable shape measurements, so we do not study them as separate structures. 
 
 Since IA is known to  depend on both the stellar and total mass \citep[e.g.,][]{tenneti-2015},   
 we have controlled for differences in the mass distributions in our samples by implementing a mass-dependent weight for each galaxy. We binned the galaxies in each sample by their total subhalo masses.   Then we took the ratio of the histogram bin heights for the \textit{Pure disc} sample to the histogram bin heights for the given sample  (because the \textit{Pure disc} sample has the narrowest distribution). These ratios   were used as weights when calculating the two-point functions using the weighted estimators in Sec.~\ref{subsec:2pcf-est}.

 Also, IA is known to depend on the satellite fraction in a given sample \citep{johnston-2019}. 
 In our samples, the satellite fractions were within 5 per cent of each other,so we did not control for this; the satellite fractions were 0.62, 0.63, 0.58 for the \textit{Elliptical}, \textit{Disc+Bulge} and \textit{Pure Disc} samples, respectively.
 Lastly, we have checked the total host (parent) halo mass distributions of all  samples after controlling for total subhalo mass and the mass distribution of the host halos is very similar, suggesting that the samples are experiencing very similar environments.  
    
\section{Results  }\label{results}
In this section, we will first describe the measured shapes of the classified galaxies and galaxy components. The misalignments of galaxy shapes and angular momenta with respect to those of the host DM halos are examined next. Third, we measure the intrinsic alignment signals of the various samples.  For large-scales we compare them with the predictions of the NLA model, while for small scales  we quantify the alignment amplitude by fitting a simple power law model.   

\subsection{Shapes of galaxies and components}
\label{shape_result}

\begin{figure*} 
\includegraphics [width=6.6in,angle=0]{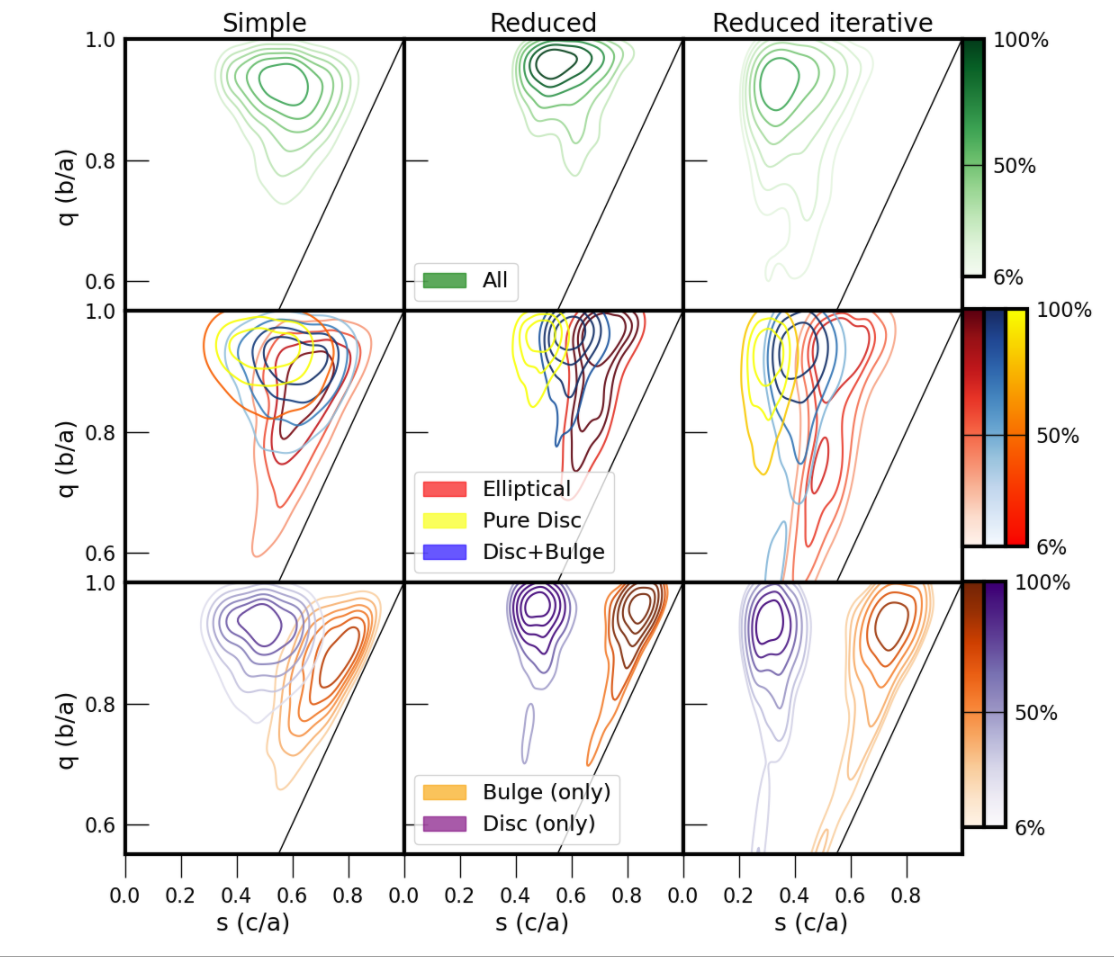}
 \caption{Comparison of shapes obtained from the three different mass quadrupole moment definitions: simple, reduced, reduced iterative ordered by columns; and rows are split by different samples as labeled in the legend, where the middle row shows three distinct samples and the bottom row shows the results for the two components of the galaxies in the {\em Disc+Bulge} sample.
 The simple method generally produces a    broader range of shapes  compared with the other two methods, stretching from $s=0.2$--$0.9$ and $q=0.6$--$1$.  
 The reduced method produces the most round shapes, as illustrated by the fact that the population in all samples is closer to the top-right edge of the plots than for the other two methods. The reduced iterative method corrects the over-rounding effect and produces a wider range of shapes than the reduced method. 
  Comparing the \textit{Disc+Bulge} sample from row 2 to row 3 (where it is split into components), the population separates out into a bimodal distribution.   
 Additionally, for row 2,  we can see that the populations with higher disc fraction are to the left, whereas populations with lower disc fraction are to the right.
 }\label{shape_ellipticity}
 \end{figure*}

  Our first test is to compare the galaxy shapes measured using the three different mass quadrupole moment definitions described in Sec.~\ref{shapes_methods}. Fig.~\ref{shape_ellipticity} shows the distribution of 3D shapes, defined as $s=c/a$ and $q=b/a$,  obtained from the three different methods of mass quadrupole moment: simple, reduced, reduced iterative. We will use it to explore the shape distributions of morphologically classified samples.  
  The reduced method produces the most round shapes, by imposing a spherically symmetric  down-weighting of particles in the galaxy outskirts, thus biasing the measurement.  
  By comparing the axis ratios for the reduced method (middle column) and reduced iterative methods (right column), we can see that the latter are systematically less round, because the iterative procedure with an elliptical weight function avoids the biases of the reduced method.
  The simple method shows overlap among the morphologically separated samples  
  in the plots  comparing axis ratios for elliptical, pure disc, and disc$+$bulge samples, while 
   
  the reduced and iterative reduced methods show similar overlap between the \textit{Disc+Bulge}  
  and \textit{Elliptical} samples. 
  In contrast, for the \textit{Disc (only)} and the \textit{Bulge (only)} samples the {\em contours} do not overlap, implying these are morphologically distinct structures within a common set of galaxies (at least when the outer particles are downweighted), however, there is still some overlap for the simple method at $\langle s \rangle \sim 0.5$. This  suggests that the disc structures may be thicker than expected from observations \citep{disc-flattening}, since for the simple method   particles are not weighted by their distance. 
  
  Interestingly, the shapes obtained by the reduced method correlate moderately (Pearson-$r$ coefficient of  +0.51 
  ) rather than strongly with the simple method shapes, which may suggest that for a significant number of galaxies  
  the outer region is quite different than the inner region.  
  The shapes from the reduced and the iterative reduced methods are strongly correlated, with a  Pearson-$r$ coefficient of 0.9. Even when corrected for the rounding of the reduced method via iteration, the shape measurements still show moderate correlation with the simple one.  
  Therefore, we will ignore the reduced method and focus on the shapes obtained by the simple and the reduced iterative methods. 

\begin{figure}
\includegraphics [width=3.0in]{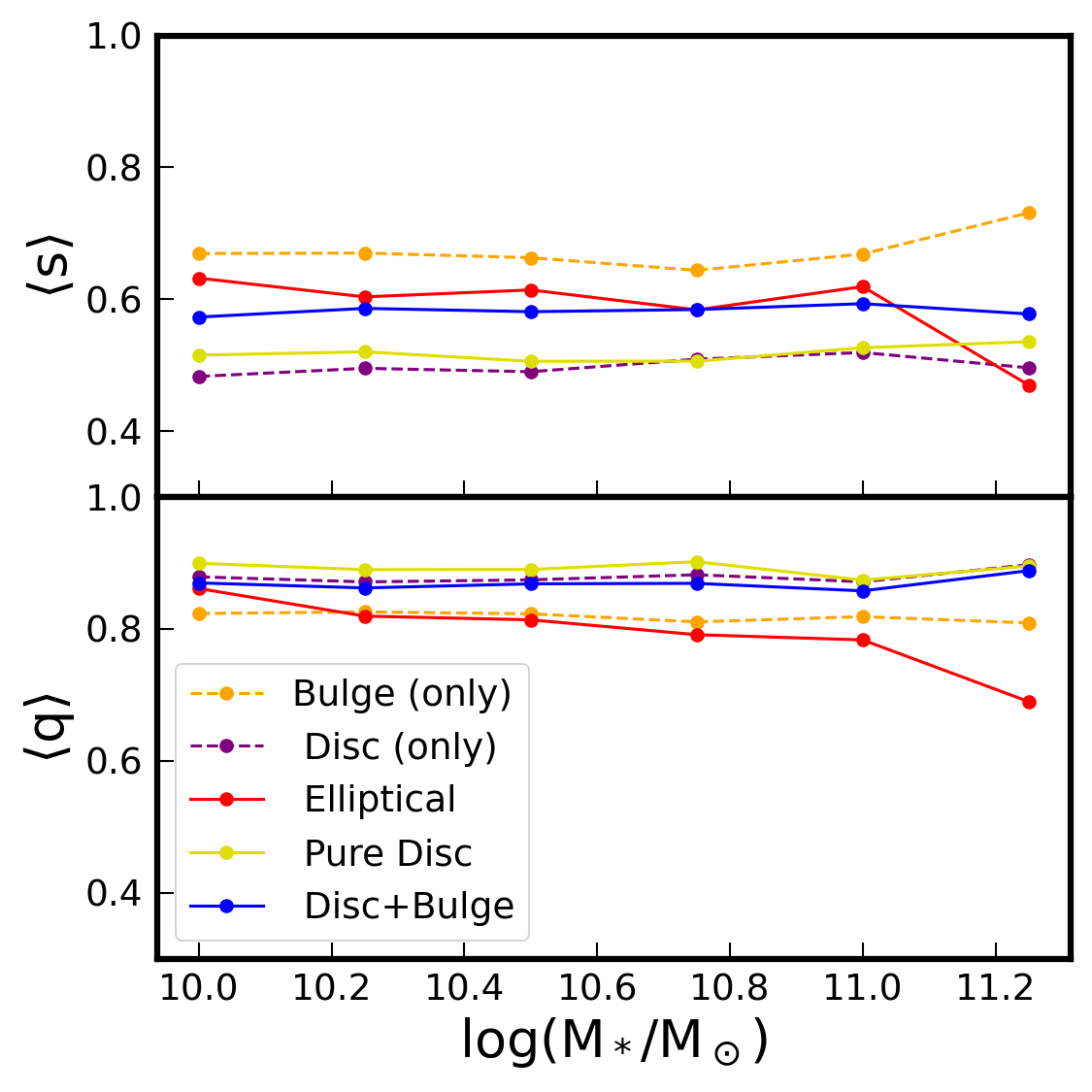}
 \caption{Comparison of average axis ratios as a function of stellar mass. For $\langle s \rangle$, the \textit{Disc (only)} populations as expected are the lowest, followed by \textit{Disc+Bulge}, and \textit{Bulge (only)} and \textit{Ellipticals} are higher. For $\langle q \rangle$, all samples are relatively flat between 0.8 and 0.9.  
 } \label{avg axis ratios}
 \end{figure}

Next, in Fig.~\ref{avg axis ratios} we investigate how the average axis ratio changes with mass. 
All samples show relatively flat trends with mass. 
For the most part, when comparing $\langle s \rangle$ and  $\langle q \rangle$, 
 
$\langle s \rangle$ appears to be lower than $\langle q\rangle$ by 20-40 per cent for a given sample, which shows that the galaxies in this simulation are generally oblate.  
As expected  for $\langle s \rangle$, the \textit{Disc (only)} sample is the thinnest (most oblate), followed by the \textit{Disc+Bulge} sample. \textit{Ellipticals} and \textit{Bulge (only)} are at a higher value, $\sim$0.7. Also, looking at plot showing dependence of $\langle s \rangle$ on mass, we see that the \textit{Pure Disc} and \textit{Disc (only)} samples are thinner by $\sim$ 20-25 percent compared to the \textit{Ellipticals} and the \textit{Bulges (only)}, which implies that the measured shapes of disc structures are relatively thick.

  \begin{figure} 
\includegraphics [width=3.3in]{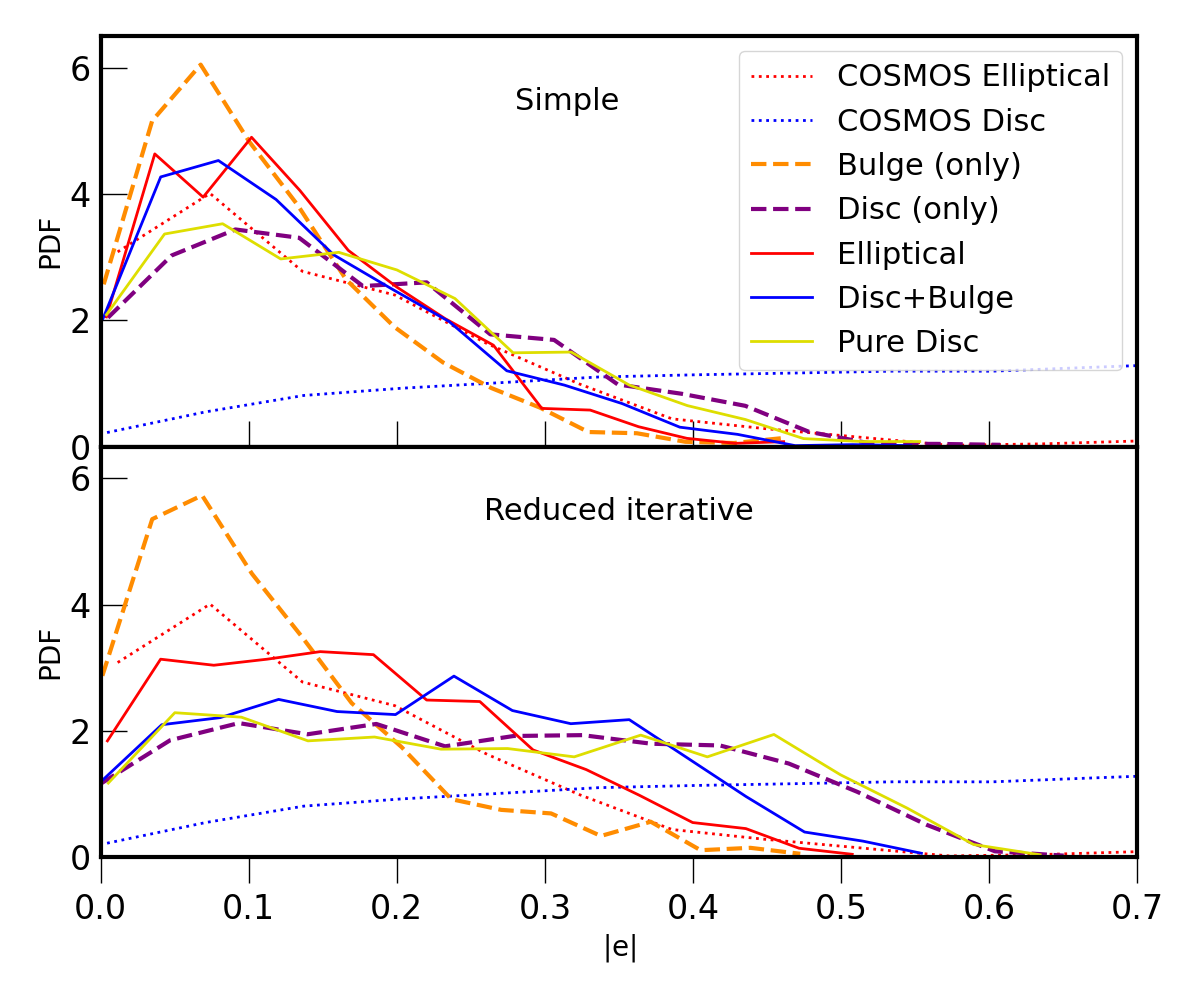}
 \caption{Distributions of the projected ellipticity magnitude $|e|$ for different simulated samples and observed values from the HST COSMOS samples from \citep{joachimi-2013}. \textit{Top panel:} Ellipticity distribution obtained using the simple mass quadrupole moment. \textit{Bottom panel:} Ellipticity distribution obtained using the reduced iterative mass quadrupole moment.  Comparing the different samples, the bulges exhibit the roundest ellipticity distribution, with a peak around 0.09. Samples with significant discs exhibit the broadest distribution. The \textit{Elliptical} and the \textit{Bulge (only)} samples have similar distributions as the \textit{COSMOS Elliptical} sample. However,  samples   with significant discs  do not agree with the \textit{COSMOS Disc} sample. For the bottom panel, \textit{Disc (only)} and \textit{Disc+Bulge} samples both show   broad distributions that   are closer to the  the \textit{COSMOS Disc} sample than the \textit{Disc (only)} and \textit{Disc+Bulge} samples from the top panel.
 }\label{ellipticities}
 \end{figure}
 
Next, we compare the 2D projected shapes, described at the end of  Section~\ref{shapes_methods}, with observed values as seen in Fig.~\ref{ellipticities}. Here, we show the ellipticity distributions of our samples and the HST \textit{COSMOS} samples from \cite{joachimi-2013}. These 2D (projected) shape distributions follow the same trend as the 3D shape distributions where \textit{Elliptical} and \textit{Bulge (only)} samples display round shapes and \textit{Bulge+Disc}, \textit{Disc (only)} and \textit{Pure Disc } samples display elongated shapes.   
Interestingly, the projected shapes obtained by the simple method show all of the samples being close to round, since the distributions have peaks at around $|e|=0.1$ but with a tail to higher values. Also, the \textit{Elliptical} and the \textit{Disc+Bulge} samples show very similar distribution, as was seen in the 3D case.  
Additionally, when compared with the \textit{COSMOS} data, the \textit{Elliptical} and the \textit{Bulge (only)} samples have similar projected shape distributions as the \textit{COSMOS Elliptical} sample; on the contrary, the \textit{Disc-related} samples do not agree with \textit{COSMOS Disc} sample, the latter of which has a very flat ellipticity distribution.

The reduced iterative 2D shape distributions in the bottom panel of Fig.~\ref{ellipticities} exhibit more variety between samples compared to those for the simple method. The \textit{Bulge (only)} sample exhibits the roundest ellipticity distribution, with a peak around 0.09 followed by a steep drop-off, followed by the \textit{Elliptical} sample. The \textit{Disc (only)} and  \textit{Disc+Bulge} samples exhibit very similar, broad  distributions that extend to $|e|\sim 0.6$; these two distributions are closer to the distribution for the \textit{COSMOS Disc} sample  than what was obtained using the simple method.   
Additionally, the \textit{COSMOS Elliptical} data is midway between the TNG-100 \textit{Bulge (only)}  and the \textit{Elliptical} samples. The disagreement between the observed and simulated shapes of the disc samples may be due to the simulation producing thick discs, as was noted in discussion of the 3D shape distributions.

\subsection{Misalignment of galaxies and components with their host DM halos}
\label{misalign_sec}
 
In order to characterize the relative orientation between the galaxy and the DM halo we focus on the relative angles between their axes, quantified using the simple quadrupole moment. For rotation-dominated structures, the minor axis should strongly coincide with the total angular momentum of the system. Since these objects are often circular in shape in the plane of the galaxy, there is also some ambiguity in determining their major and the intermediate axis, causing a degeneracy in these axes. 
 
\begin{figure*}
\includegraphics [width=6.6in,angle=0]{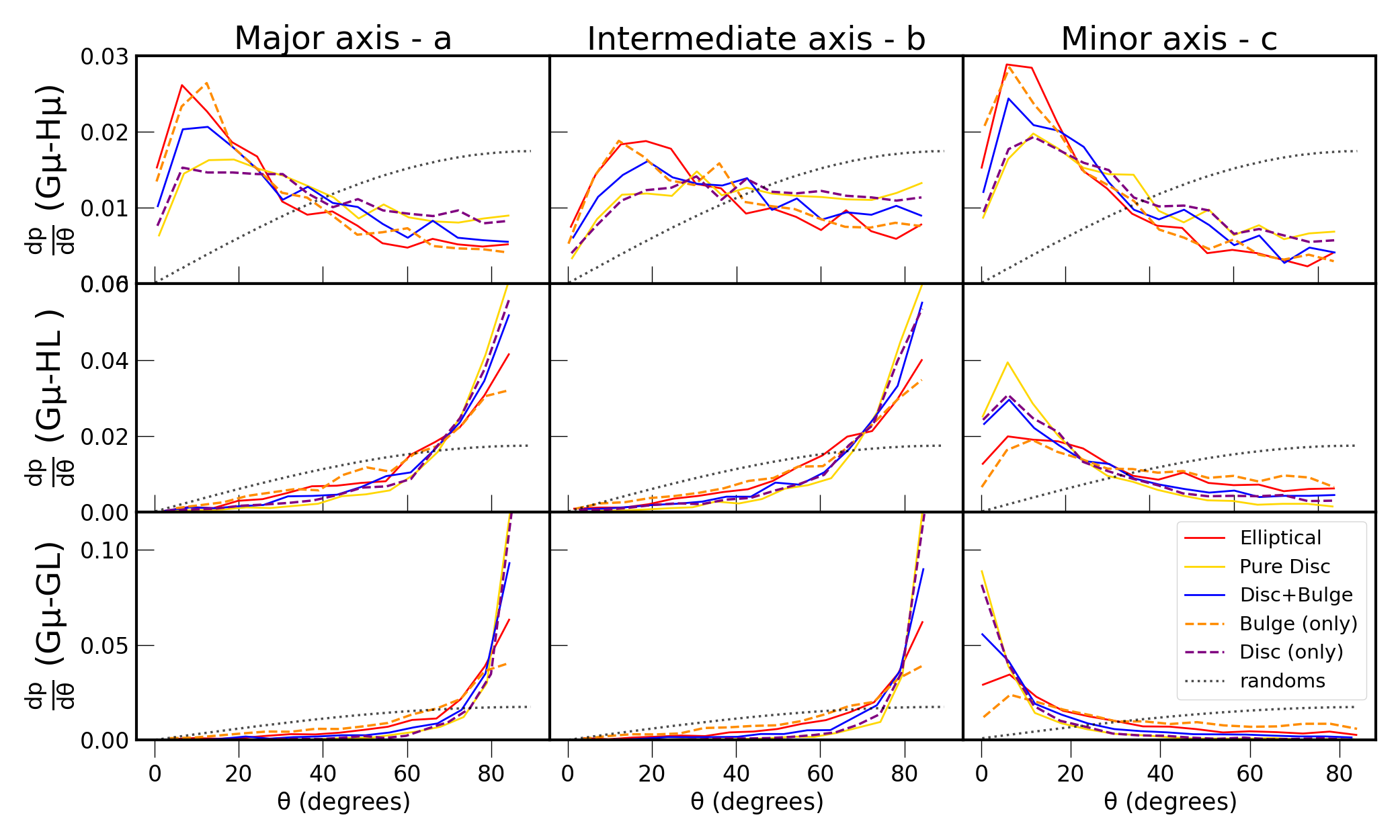}
 \caption{Distribution of misalignment angles of various shape axes (measured using the simple quadrupole moment) and total angular momenta. [G = galaxy, $\mathrm{\mu}$ = axis $\in (a,b,c)$, H = DM halo, L = total angular momentum, e.g. G$\mathrm{\mu}$-HL means misalignment angle between the galaxy axis and the DM halo angular momentum]. The grey dotted line in all panels represents random alignment. The minor axis of the galaxy for the most part aligns well with the minor axis of the DM halo and the two angular momentum vectors, especially for the  \textit{Disc+Bulge} and \textit{Disc (only)} samples. Conversely, the major and intermediate axes are mostly perpendicular to the two $L$ vectors. The galaxy and DM halo major and intermediate axes shows sign of preferred alignment compared to a random distribution for all samples. 
 } \label{misalign}
 \end{figure*}

In Fig.~\ref{misalign} we show the distribution of various misalignment angles of different galaxy types and components and in Table ~\ref{misalign_table} we tabulate the median, the 16th and the 84th percentiles for the top-left, middle-right, bottom-right panels in Fig. 4. 
In all panels,  the grey dotted lines denote the expected distribution for purely random orientations; as shown, all galaxy types and components have alignments that are not random with respect to the halo shape, and the galaxy and halo angular momentum.
 
In the plots showing the alignment angles between the galaxy axis and halo axis in the top row, \textit{Ellipticals} show stronger alignment  compared with the other samples, consistent with the physical picture that elliptical galaxies tend to be stretched out along the direction of the tidal fields. Interestingly, the \textit{Bulge (only)} sample follows a very similar trend   as the \textit{Ellipticals} suggesting that bulges also tend to be stretched out by the tidal fields. 
Comparing the alignments of the intermediate and major axes, we see that the alignments are less significant for the intermediate axes for the \textit{Ellipticals} and \textit {Bulges (only)}, however for \textit{Pure Disc} and \textit{Disc (only)} samples the curves look very similar, since  there is some ambiguity in determining the major and intermediate axes for oblate objects. For all samples, the minor axes of the galaxy and DM halo shape exhibit the strongest alignment compared to the other two axes, again with \textit{Ellipticals} and \textit{Bulges (only)}  being the most strongly aligned (with median $\theta = 20.34, 21.77$ respectively), followed by the disc-related structures. 
 
Next, we inspect how the galaxy axis aligns with respect to the halo angular momentum vector as shown in the second row of Fig.~\ref{misalign}. As expected, the galaxy major and intermediate axes exhibit a tendency to anti-align with the halo angular momentum vector (i.e., systematically less aligned than a random distribution) for all samples. The galaxy minor axis, as expected, has a strong alignment with the halo angular momentum for the \textit{Pure Disc} , \textit{Disc+Bulge} and \textit{Disc (only)} cases (with median $\theta = 15.99, 21.13, 18.97$, respectively) , suggesting a strong torquing mechanism between the DM halos and the resident galaxies.  
In contrast, the \textit{Bulge (only)} and \textit{Elliptical} samples show weaker alignments  with the halo angular momentum (with median $\theta = 36.59, 28.87$, respectively), implying a weaker torquing mechanism from the DM halo to the resident galaxies.  
 
 Lastly, the bottom row of Fig.~\ref{misalign} shows that there is a strong alignment between the minor axis of disc-like samples and the galaxy's total angular momentum, with \textit{Pure Disc} , \textit{Disc+Bulge} and \textit{Disc (only)} showing low median $\theta$ of 5.51, 9.29, 6.09, respectively . While the \textit{Bulge (only)} and \textit{Elliptical} samples exhibit lower alignment with higher median $\theta$  of 29.41 and 17.16, respectively.  
 Comparing how the galaxy shape axes are aligned with the halo and galaxy angular momentum vectors, the galaxy shape minor axis is more aligned, and the major and the intermediate shape axes are  more anti-aligned   with the galaxy angular momentum vector.  
 Also, for these angular momentum vector alignments with the galaxy axes, the ambiguity in determining the major and intermediate axes is not relevant, since we are primarily interested in the alignment of the minor axes with the angular momentum vectors. 
  
  Overall, all samples exhibit a preferred alignment of their galaxy and DM halo shape axes, with the  \textit{Ellipticals} and \textit{Bulge (only)} samples showing a stronger alignment compared to the other samples. Also, all samples tend to align their galaxy minor axes with the galaxy angular momentum and to a lesser degree with the DM halo angular momentum, with the \textit{Pure Disc}, \textit{Disc+Bulge} and \textit{Disc (only)} samples exhibiting stronger alignment than the \textit{Elliptical} and \textit{Bulge (only)} samples.

\begin{figure*}
\includegraphics [width=6.6in,angle=0]{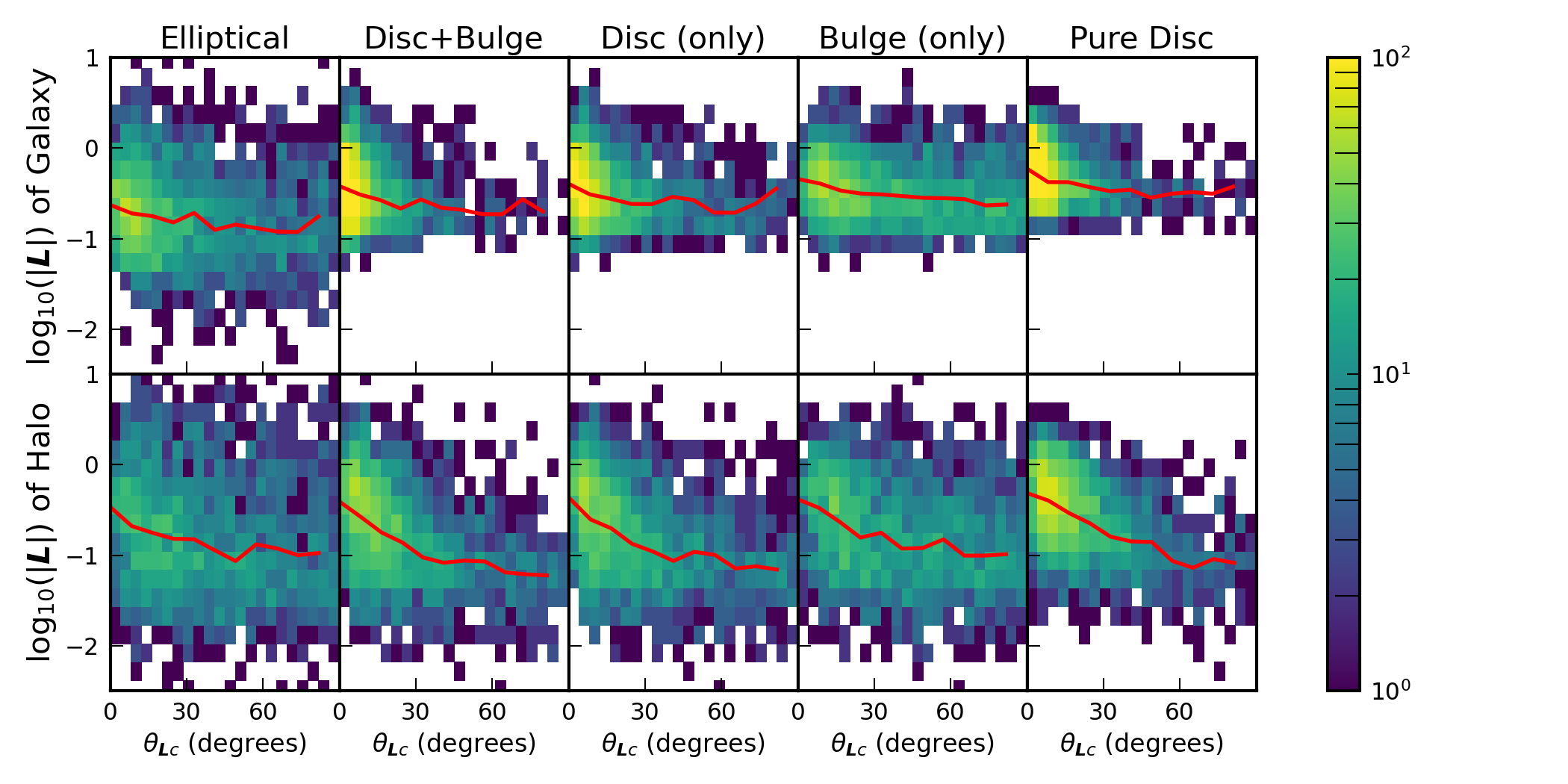}
 \caption{\textit{Top row:} Distribution of   the magnitude of the total angular momentum of galaxies vs.\ the misalignment angle with the minor axis of the galaxy (measured using the simple quadrupole moment), where $\mathbf{L}$ is in units of $\mathrm{Mpc \, km}\sqrt{a}/s \, M_\odot$. \textit{Bottom row:} The same, for the magnitude of the total angular momentum of halos.  
 The color bar indicates the number of galaxies in each bin, while the red line indicates the median for each panel at a given angle. When contrasting the top and the bottom row, the galaxy angular momentum is more aligned with the minor axis of the galaxy shape than the halo's angular momentum for each sample.   
 For the bottom row, the \textit{Disc+Bulge}, \textit{Disc (only)} and the \textit{Pure Disc} samples show high concentration of population near 0 degrees, whereas the \textit{Elliptical} and the \textit{Bulge (only)} samples show smooth distribution with no highly concentrated areas. Overall, the decreasing trend indicated by the red line shows that the alignment is not dominated by low magnitude angular momentum vectors.  
 }\label{c_vs_L}
 \end{figure*}

Further, to test the dependence of the alignments with the total angular momentum vectors shown in Fig.~\ref{misalign} on the angular momentum magnitude,   we explore the relationship between the alignment angle $\theta_{\bm{L}c}$ with the magnitude of the angular momentum vector $|\vec{L}|$ of the galaxy and the DM halo in Fig.~\ref{c_vs_L}.  When comparing different samples, the \textit{Disc+Bulge}, \textit{Disc (only)} and \textit{Pure Disc} samples show a high concentration of probability density near 0 degrees, indicating strong alignment,  
whereas the \textit{Elliptical} and the \textit{Bulge (only)} samples have a smooth distribution with no highly concentrated areas.   
All panels in this plot show  that a low  angular momentum magnitude $|\vec{L}|$  leads to significant misalignment, consistent with the physical picture that significant angular momentum is required in order for it to drive   galaxy alignments. 
 
Galaxy-halo axis misalignment has been studied in various cosmological hydrodynamical simulations:  in MassiveBlack-II  
and Illustris \citep{tenneti-disc-ellip}, in Horizon-AGN \citep{horizon-misal}, and  in EAGLE \citep{eagle-misal}. Both \cite{tenneti-disc-ellip} and \cite{horizon-misal} found that elliptical galaxies  in the  simulations have stronger   
alignments with their host halos than do disc galaxies, similar to what we have found in TNG100. In contrast, \cite{eagle-misal} reports that elliptical galaxies are more misaligned with their host halos compared to disc galaxies.  
We note that each of these studies used slightly different dynamical methods to identify the galaxy morphology, and the studies did not control for intrinsic alignment trends with mass, though they explored trends in mass by breaking their samples into mass bins.
Furthermore, this study and those cited above  
find that galaxy shapes exhibit coherent alignments with respect to host halo shapes, and that the elliptical versus disc galaxy distinction is necessary for accurate models of   misalignment angles.  
As we will see in Sec.~\ref{NLA_section}, galaxy samples that are more aligned with their host DM halos exhibit higher large-scale galaxy alignment correlations, which we can quantify using the NLA model amplitude.

\begin{table*}
%\centering
 
 \begin{tabular}{c||c c c c c c c||} 
 \hline
       & Elliptical & Disc+Bulge & Bulge (only) & Disc (only) &Pure Disc\\ [0.5ex] 
 \hline\hline
 Gc-Hc (top-left in Fig.~\ref{misalign}) 
        & $20.34^{+49.96}_{-8.10}$ & 
         $26.79^{+57.78}_{-9.99}$ &
         $21.77^{+50.81}_{-8.50}$ & 
         $31.92^{+63.68}_{-12.36}$ &
         $32.97^{+66.11}_{-12.80}$

 \\ 
 \hline
  Gc-HL (middle-right in Fig.~\ref{misalign})  
        & $28.87^{+64.01}_{-9.69}$ & 
         $21.13^{+53.95}_{-7.27}$ &
         $36.59^{+70.68}_{-13.23}$ & 
         $18.97^{+47.42}_{-6.77}$ &
         $15.99^{+39.39}_{-6.42}$  
 \\ 
 \hline
  Gc-GL (bottom-right in Fig.~\ref{misalign})  
        & $17.16^{+51.77}_{-5.74}$ & 
         $9.29^{+32.04}_{-3.23}$ &
         $29.41^{+68.79}_{-9.94}$ & 
         $6.09^{+18.83}_{-2.27}$ &
         $5.51^{+17.60}_{-1.96}$ \\
 \hline

\end{tabular}
\caption{   Median and 16th and 84th percentiles of misalignment angles of galaxy-halo major axes, and galaxy minor axes against total angular momenta of galaxy and halo. This Table is intended to supplement Fig. ~\ref{misalign} by providing quantitative information on the plotted distributions. 
}
\label{misalign_table}
\end{table*}

\subsection{Alignment measurements from two-point functions  }
\label{NLA_section}
 
In this section we present and discuss the results from the  joint fits to the NLA model that were performed using the $w_{gg}$, $w_{g+}$, $w_{++}$ data vectors.
We found that $w_{g+}$ for the reduced  iterative method is consistent with zero, similar to the findings of a previous study done by  \citet{chisari-horizon-ia} using the reduced method. This may be due to particles in the outer regions of galaxies being affected more by the tidal forces from the large-scale structure compared to the particles in the inner regions \citep{Singh-2016,horizon-misal}.  
Hence in this section we focus on the results obtained using the simple mass quadrupole moment.  
Since the NLA model is known to break down at small scales ($r_p$ less than a few Mpc),  
our fits to the NLA model only include $r_p> 4$~Mpc \citep{bridle-king}.

\begin{figure*}
   \hspace*{-0cm}
\includegraphics [width=7.0in,angle=0]{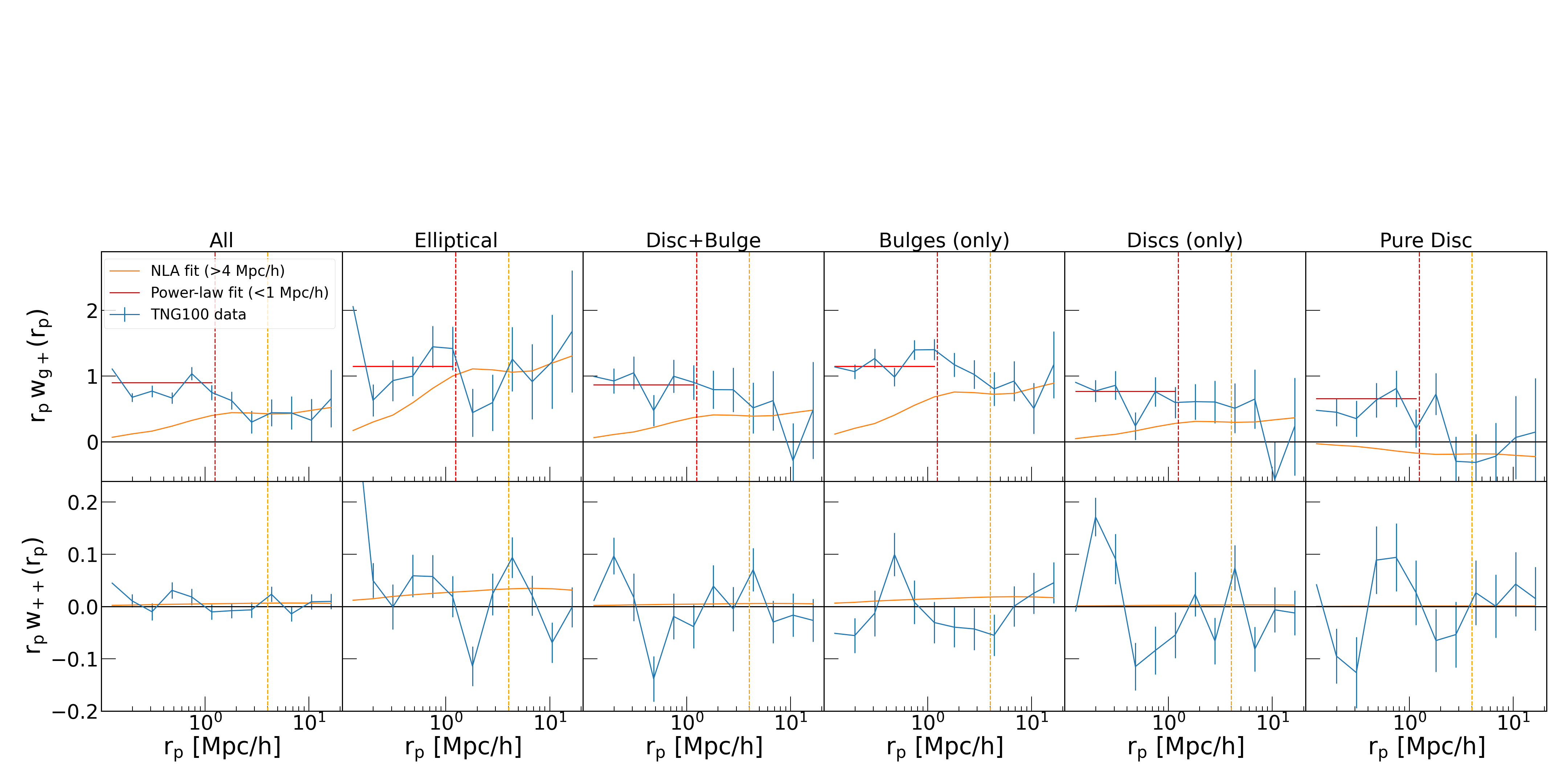}
 \caption{Measured quantities $w_{g+}$ (top row) and $w_{++}$ (bottom row) in blue for 6 different dynamically classified/decomposed mass-controlled samples  as indicated in the column titles. Data on scales below  $r_p < 4$~Mpc (indicated by the orange dashed vertical line) were excluded from the NLA fits.  For each sample (i.e., for the data in a given column),   $w_{g+}$, $w_{++}$ along with density tracer $w_{gg}$ ($w_{gg}$ curve not shown here, since it was the same for all samples)  were fit simultaneously  
 to the NLA model using a Markov Chain Monte Carlo method in the range $r_p < 4$~Mpc; the best-fitting parameters are shown in row 1 of Table~\ref{nla_table}.  
 In order to quantify the small scale alignments, we fit a power-law function (Eq.~\ref{power-law-eq})  
 for $r_p < 1$~Mpc (indicated by the red dashed vertical line).
 The power-law fit parameters are presented in Table~\ref{power_law_table}. The best fitting curves for the NLA model are shown in orange  and for the power-law model in red. We show the  
 extrapolation of the best-fitting NLA model to small scales, to justify our choice of $r_p$ cut. 
 }\label{nla_fit}
 \end{figure*}

The measured values and the best-fitting curves for   $w_{g+}$, $w_{++}$ are shown in Fig.~\ref{nla_fit}.   We omitted $w_{gg}$ for the density tracer sample since it was the same in all the fits.  
In general, the NLA model fits the measured values from the simulation well on large scales. As in previous studies \citep{mandelbaum-2006, singh-2015},  
the  $w_{++}$ curves are too noisy to carry out meaningful model fits  
and were consistent with zero;  
we only include them in the joint fits for the NLA model at large-scales rather than trying to model them separately. Also, from the extrapolation of the best-fitting curves, one can see that the NLA model severely underestimates $w_{g+}$ at small scales, confirming our previous assumption that the model will fail on smaller scales.  In all of the samples the galaxy bias parameter from the joint fits was around 1; the bias parameter being the same across samples is expected, since we have used the same density tracer sample in all cases.

\begin{figure}
\includegraphics [width=3.0in,angle=0]{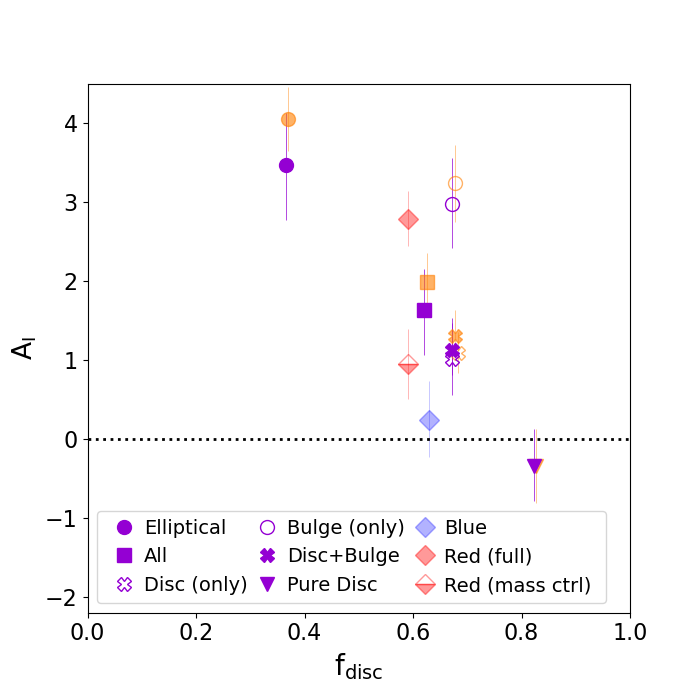}
 \caption{Dependence of the alignment strength parameter $A_I$ on $f_\text{disc}$. The purple points are the main mass-controlled samples presented in the top row of Table~\ref{nla_table}. The orange points are the non-mass-controlled samples (i.e., the full samples from the simulation, shown in the second row of Table~\ref{nla_table}). In both cases, the alignment strength exhibits a decreasing trend with $f_\text{disc}$. The purple points are slightly below the orange points, indicating that there is some dependence on mass (with higher mass resulting in greater alignments), however the overall trend with $f_\text{disc}$ is the same.
 Additionally we have plotted the color-split samples with the diamond-shaped points; for the \textit{Red} sample, the half-filled diamond represents the mass-controlled sample and the full diamond represents the full sample. Purple points were arbitrarily shifted by 1\% to the left for visual clarity. 
 }\label{AI_fdisk}
 \end{figure}

In row 1 of Table~\ref{nla_table} we show the best fitting parameters of the NLA model for our mass-controlled samples.  These results are visualized in Fig.~\ref{AI_fdisk} with purple points showing $A_I$ versus $f_\text{disc}$. As expected, the \textit{Elliptical} galaxies have the highest $A_I$ of $3.47^{+0.57}_{-0.57}$, whereas the Pure Disc population has an $A_I$ value consistent with zero.  In addition, when the two-component Disc+Bulge galaxies are split into their separate components, the \textit{Bulge (only)} sample exhibits a high $A_I$ signal comparable to that of \textit{Ellipticals} galaxies, at $2.98^{+0.36}_{-0.37}$.  In contrast,  the \textit{Disc (only)} component has a lower $A_I$ value of $1.02^{+0.34}_{-0.36}$, still different than that for the \textit{Pure Disc} sample, corresponding to a marginal detection $2.5\sigma$ from zero.  
The two-component galaxies as a whole have a weak signal at $A_I=1.13^{+0.37}_{-0.35}$ that is consistent with that for the \textit{Disc (only)} sample; this is understandable since the shape of those galaxies is dominated by the disc structure when using the simple shape estimator.   
In Fig.~\ref{AI_fdisk} we see a clear downward trend of $A_I$ with $f_\text{disc}$, and with the \textit{Bulge (Only)} sample showing consistent alignment as the \textit{Elliptical} sample.   
Compared with previous IA studies using hydrodynamical simulations, our results for the whole non-mass controlled sample agree with \cite{samuroff-2019}, who report $A_I=1.71^{+0.17}_{-0.17}$ and $b_g=1.11^{+0.07}_{-0.07}$, though it should be noted that they used the larger volume TNG300 and a lower mass-cut of $\log_{10}(M_*/M_\odot) \approx 9.37 $ on their sample.  
 
We have investigated how the intrinsic alignment results for mass-controlled   samples compare with the full (i.e. non mass-controlled samples) samples. We did the same analysis for the full samples; row 2 of Table~\ref{nla_table} shows the resulting NLA fit parameters. As expected given the higher average mass, the full samples shows higher $A_I$ value by $\sim$5-15\%  
compared to the mass controlled samples.  
Nonetheless, the trends with $f_\text{disc}$ are very similar in both cases as illustrated by Fig.~\ref{AI_fdisk}: both the mass-controlled and full samples exhibit a decreasing trend in $A_I$ with increasing $f_\text{disc}$.   

In observational tests of intrinsic alignments, the galaxy samples are usually split by color, since morphological information is  typically  unavailable and color serves as a proxy for morphology.   
Hence, in this study we also measure the alignments for samples split by color to provide better comparison with observational studies. We split the full sample in the $g-r$ color versus $r$  magnitude 
 
plane using the same method as in \cite{samuroff-2020}.  The results of NLA fits to the alignment signals are summarized in Table~\ref{color_table}. For the \textit{Red} sample we present both the mass-controlled   and the full sample; however the \textit{Blue} sample was not mass controlled because of its very narrow mass range. Here, we should note that the simulation produces a slightly higher number of red disc galaxies compared to observations, which complicates the interpretation of these results  \citep{tng-bimodal,gal_decomp}.  
The \textit{Blue} sample has an IA signal that is consistent with zero, similar to previous studies in real data such as \cite{johnston-2019,mandelbaum-2011}. The \textit{Red} sample has alignment amplitude $A_I=2.79^{+0.50}_{-0.35}$,  consistent with the $A_I =3.63^{+0.79}_{-0.79}$ of the GAMA Z1R sample in \citep{johnston-2019}.  The two samples have comparable mean luminosities of $\langle L \rangle/L_0=0.45 $ and $0.50$, respectively, where $L_0$ is the pivot luminosity corresponding to $M_r=-22$.  
One interesting feature of this result is that the morphologically-selected samples, \textit{Ellipticals} and \textit{Bulge (only)}, both exhibit a higher $A_I$ signal compared to the color-selected \textit{Red} sample.  
The mass-controlled \textit{Elliptical} sample shows almost thrice as strong IA signal compared to the mass-controlled \textit{Red} sample. 
 
 Compared with observational results from \cite{fortuna-kids-lrg-2021}, their \textit{dense red} sample with average luminosity of  $\langle L \rangle/L_0=0.38$ has the same average luminosity as our \textit{Elliptical}  sample, and has $A_I=3.69^{+0.66}_{-0.65}$, which agrees with both our mass-controlled and full elliptical sample. Also their \textit{L1} sample, which is a subset of their luminous red galaxy sample, with  $\langle L \rangle/L_0=0.46$ and $A_I=1.80^{+0.96}_{-0.95}$, has consistent alignment amplitude with both the mass-controlled and the full \textit{Red} samples  within the error bar.  
 
\begin{table*}
%\centering
 
 \begin{tabular}{c||c c c c c c c||} 
 \hline
     &All & Elliptical & Disc+Bulge & Bulge (only) & Disc (only) &Pure Disc\\ [0.5ex] 
 \hline\hline
 
 mass controlled  $A_I$ & $1.64^{+0.28}_{-0.28}$ &
         $3.47^{+0.57}_{-0.57}$ & 
         $1.13^{+0.37}_{-0.35}$ &
         $2.98^{+0.36}_{-0.37}$ & 
         $1.02^{+0.34}_{-0.36}$ &
         $-0.34^{+0.35}_{-0.36}$

 \\ 
 \hline
  non-mass controlled $A_I$ & $1.99^{+0.19}_{-0.19}$ &
         $4.03^{+0.36}_{-0.34}$ & 
         $1.31^{+0.22}_{-0.20}$ &
         $3.24^{+0.22}_{-0.25}$ & 
         $1.09^{+0.27}_{-0.29}$ &
         $-0.34^{+0.35}_{-0.36}$ \\
 
 \hline

\end{tabular}
\caption{Best fitting parameters of the NLA model using $w_{gg}$, $w_{g+}$, $w_{++}$ data points simultaneously, for $r_p > 4$~ Mpc.  
As expected, bulge-dominated galaxies exhibit a higher $A_I$. In addition, when disc-dominated galaxies are split by their components, \textit{Bulge (only)} shows high $A_I$ signal comparable to that of \textit{Elliptical} galaxies, followed by \textit{Disc+Bulge} and \textit{Disc (only)} samples.  The \textit{Pure Disc} sample shows an alignment signal consistent with zero.   
For the non-mass controlled samples (i.e., full samples without any downselection to match the mass distributions), the alignment signals are typically stronger by 10-20\% compared to the mass-controlled samples, but the trends with morphological subsample are very similar to those for the mass-controlled samples, see Fig.~\ref{AI_fdisk} for direct illustration.
}
\label{nla_table}
\end{table*}

 In \S~\ref{misalign_sec}, we saw that the \textit{Ellipticals} and \textit{Bulge (only)} samples were more aligned with their host DM halo compared to the \textit{Disc (only)} and \textit{Pure Disc}   samples.  That finding, taken together with the results of this section, suggests that the local alignment of galaxy and DM halo orientations correlates with the large-scale alignment of galaxies \citep{heymans-2006}.  
In order to quantify the small-scale intrinsic alignment amplitude, we employ an empirical power-law model of the form: 
\begin{equation}
    w_{g+}(r_p) = w^0_{g+} \Bigg( \frac{r_p}{ 1 \,h^{-1}\mathrm{ Mpc}} \Bigg)^{-1},
    \label{power-law-eq}
\end{equation}
fitting the measured data below 1 Mpc.   
In Table~\ref{power_law_table} we present the best fitting parameters of the power law model.  
All the fits for $w_{g+}$ were done independently, as opposed to the NLA model where $w_{gg}$, $w_{g+}$, and $w_{++}$ were fit jointly, %  
for $r_p<$1~Mpc.
 
Comparing with other work, we note that \cite{mandelbaum-2006} (which is from SDSS) and \cite{chisari-horizon-ia} (which is from the Horizon-AGN simulation) report  values of 
$w_{g+}^0 = 0.10^{+0.07}_{-0.07}$ and $0.13^{+0.03}_{-0.03}$, respectively, for their whole sample. Both of these are  about a factor of 7 smaller than what we have measured; however, they had their power-law index as an additional fit parameter, unlike ours, which was fixed at -1, and they report   $-0.59^{+0.65}_{-2.30}$ and $-0.75^{+0.25}_{-0.25}$ 
for their power-law index. Additionally, we expect the amplitude of \cite{mandelbaum-2006} and \cite{chisari-horizon-ia} to be lower since they used lower  mass cuts of $   \log_{10}(M_*/M_\odot) \sim 9 $ and $\log_{10}(M_*/M_\odot) \sim  8 $, respectively.
Also, non-linear bias was not modeled, so it is hard to explain the disagreement quantitatively.
Thus, our fixed power-index of -1 is within the range of both of these studies, despite the strong disagreement in the amplitude $w_{g+}^0$.

\begin{table}\label{color-table}
\centering
 \begin{tabular}{||c c c c ||} 
 \hline
   &Red (full sample) & Blue &Red  (mass controlled) \\ [0.5ex] 
 \hline\hline
 $A_I$   & $2.79^{+0.42}_{-0.43}$ & $0.13^{+0.35}_{-0.34 }$  & $0.95^{+0.41}_{-0.45}$ \\ 
 \hline
 
\end{tabular}
\caption{Best fitting parameters of NLA model when we split the whole sample by color rather than morphology. The two color bins were split in the $g-r$ color versus $r$ magnitude plane.  The full \textit{Red} sample has an alignment amplitude $A_I$ of $2.79^{+0.42}_{-0.43}$,  
whereas the \textit{Blue} sample has $A_I$ consistent with zero. When mass-controlled, the \textit{Red} sample's alignment strength decreases by more than a factor of two. The mean $f_\text{disc}$ values were  0.63, 0.61  and 0.59 for the \textit{Blue}, mass-controlled \textit{Red} and full \textit{Red} samples, respectively. The \textit{Blue} sample was not mass controlled, since the \textit{Blue} sample's mass range was very narrow. 
}
\label{color_table}
\end{table}

\begin{table*}
\centering
 \begin{tabular}{||c c c c c c c||} 
 \hline
   &All & Elliptical & Disc+Bulge & Bulge (only) & Disc (only) &Pure Disc\\ [0.5ex] 
 \hline\hline

$w_{g+}^0$ & $0.79^{+0.07}_{-0.06}$ & 
             $1.15^{+0.11}_{-0.11}$ &
             $0.67^{+0.16}_{-0.15}$ &
             $1.25^{+0.11}_{-0.11}$ &
             $0.59^{+0.17}_{-0.16}$ &
             $0.60^{+0.15}_{-0.15}$ \\
 \hline
 
\end{tabular}
\caption{Best fitting amplitude parameter $w_{g+}^0$ from the power law model fits to $w_{g+}$ for $r_p < 1$~Mpc.  These amplitudes show a very similar trend as the NLA fit amplitudes on large scales: the \textit{Elliptical} and \textit{Bulge (only)} samples have higher amplitudes, followed by the \textit{Disc+Bulge} and \textit{Pure Disc} samples, with the \textit{Disc (only)}  sample having the weakest amplitudes.  
}
\label{power_law_table}
\end{table*}

\section{Conclusions}\label{conc}

 In this work, we have investigated the intrinsic alignments of dynamically decomposed/classified galaxies using the TNG100 hydrodynamical simulation from the IllustrisTNG simulation suite.   
 As a first test, we measured the 2D and 3D shapes of simulated galaxies divided into samples based on their morphologies. We found that the distributions of the absolute value of the ellipticities for the \textit{Elliptical} and \textit{Bulge (only)} samples compare well with those from real COSMOS data. However, the shapes of the disc populations differ from those of the COSMOS dataset, overall showing more round shapes with the absolute magnitude of the ellipticity distribution always being lower than 0.6 compared to that for the COSMOS dataset, which stretches far beyond 0.6.  
 Second, we have investigated how the shapes of the different galaxy components and types align with the shapes of their host halos and with their total angular momentum vectors. Our investigation shows that the \textit{Elliptical} and \textit{Bulge (only)} samples are more strongly aligned with their host DM halos compared to the disc-related samples, in agreement with previous studies. Additionally, when measuring the galaxy shape alignments with the total angular momentum vectors of the DM halo and of the galaxy, all sample show preferred alignment, with the angular momentum-dominated \textit{Pure Disc}, \textit{Disc+Bulge} and \textit{Disc (only)} samples showing  a higher degree of alignment compared to the \textit{Bulge (only)} and the \textit{Elliptical} samples.   
 
 Third, we have measured two-point correlation functions and fit them using the NLA model for scales $r_p> 4$~Mpc and a power-law model for scales  $< 1$~Mpc.  
 As expected, the \textit{Elliptical} sample exhibits the highest value of $A_I=3.47^{+0.57}_{-0.57}$, closely followed by  the \textit{Bulge (only)} sample with  $A_I=2.98^{+0.36}_{-0.37}$. These two samples have statistically consistent alignments.  In contrast, the \textit{Disc+Bulge} and \textit{Disc (only)} samples exhibit a relatively low alignment of $A_I = 1.13^{+0.37}_{-0.35}$ and $A_I = 1.02^{+0.34}_{-0.36}$,respectively.  
 Lastly, the \textit{Pure Disc} sample shows an alignment consistent with zero. 
In order to account for potential trends in intrinsic alignments with mass, we have implemented mass-dependent weights in our two-point function calculations.  Therefore, our results cannot be explained by different host halo masses for morphologically-divided samples, but rather must be explained by different alignments at fixed mass.

We have also considered samples divided based on their rest-frame color.  The full unweighted \textit{Red} sample has a  value of $A_I=2.79^{+0.42}_{-0.43}$, which is smaller than the results for the \textit{Elliptical} and \textit{Bulge (only)} samples and even smaller when mass-controlled at $A_I=0.95^{+0.41}_{-0.45}$. Though it should be noted that the simulation produces higher number of red galaxies compared to observations which makes interpretability of the color classified results complicated. In contrast, the \textit{Blue} sample (which was not mass controlled due to its very narrow mass range) has an IA signal consistent with zero.   
 
 Modeling and accounting for intrinsic alignments will be essential for the next generation of cosmological surveys such as LSST, Roman and Euclid in order to achieve unbiased cosmological parameter estimates given their unprecedented statistical precision.  
 In this study, we had tried to provide better understanding of IA based on the stellar dynamics of galaxies. We have shown that IA decreases as angular momentum starts to dominate the stellar dynamics: large-scale IA becomes undetectable in our sample  
 when angular momentum almost completely dominates the stellar dynamics of galaxies. Dispersion-dominated systems like \textit{Ellipticals} and \textit{Bulges} exhibit a preferential alignment towards matter densities and are generally more strongly aligned with the shape of their host DM halos. This finding that the \textit{Elliptical} and \textit{Bulge} samples both have a strong IA signal on cosmological scales provides further evidence for the Elliptical-Bulge likeness (or the hypothesis that bulges are just scaled down ellipticals). In future work, it would be useful to pursue a similar study in higher redshifts, in order to track the evolution of IA in time, akin to the analysis of \cite{bhowmick}, but with morphological separation.  
 Another important aspect is baryonic physics effect on IA; it would be valuable in the future to perform a similar analysis on the gas structure of galaxies and investigate the effects of star formation and active galactic nuclei on IA, similar to \cite{tenneti-disc-ellip, samuroff-2020}; these papers studied IA in different simulations with different baryonic physics implementations.

 Our findings indicate that intrinsic alignments are more complex than is implied by the conventional Red-Blue (Elliptical-Disc) separation, with  the underlying stellar dynamics playing a significant role in determining the alignments. For current and future weak lensing studies,  IA models  that connect to the underlying disc fraction and relate to individual galaxy components may improve the robustness and flexibility of the IA models. 
 
 \section*{Data Availability}
 
 The data used in this paper is publicly available.
The IllustrisTNG data can be obtained through the website at
\url{https://www.tng-project.org/data/}. The catalog data with morphological decompositions of galaxies is available at \url{https://github.com/McWilliamsCenter/gal_decomp_paper}

\section*{Acknowledgements}

We thank Simon Samuroff, Duncan Campbell, Markus Michael Rau, Scott Dodelson for useful discussion that informed the direction of this work. We thank the anonymous referee for valuable feedback that helped improve the final version of the paper . We, also thank Benjamin Joachimi for providing us the COSMOS ellipticity dataset. This work was supported in part by the National Science Foundation, NSF AST-1716131 and by a grant from the Simons Foundation (Simons Investigator in Astrophysics, Award ID 620789). SS is supported by a McWilliams postdoctoral fellowship at Carnegie Mellon University.

\bibliographystyle{mnras}
\bibliography{example}

\bsp	% typesetting comment
\label{lastpage}
\end{document}